\documentclass{osa-article}

\journal{oe}

\articletype{Research Article}

\begin{document}

\title{Improved comb and dual-comb operation of terahertz quantum cascade lasers utilizing a symmetric thermal dissipation}

\author{Chenjie Wang,\authormark{1,2} Ziping Li,\authormark{1} Xiaoyu Liao,\authormark{1,2} Wen Guan,\authormark{1,3} Xuhong Ma,\authormark{1,2} Kang Zhou,\authormark{1,2} J. C. Cao,\authormark{1,2} and Hua Li\authormark{1,2,*}}

\address{\authormark{1}Key Laboratory of Terahertz Solid State Technology, Shanghai Institute of Microsystem and Information Technology, Chinese Academy of Sciences, 865 Changning road, Shanghai 200050, China\\
	\authormark{2}Center of Materials Science and Optoelectronics Engineering, University of Chinese Academy of Sciences, Beijing 100049, China.\\
	\authormark{3}School of Information Science and Technology, ShanghaiTech University, 393 Middle Huaxia Road, Shanghai 201210, China.}

\email{\authormark{*}hua.li@mail.sim.ac.cn} 



\begin{abstract}
In the terahertz frequency range, the quantum cascade laser (QCL) is a suitable platform for the frequency comb and dual-comb operation. Improved comb performances have been always much in demand. In this work, by employing a symmetric thermal dissipation scheme, we report an improved frequency comb and dual-comb operation of terahertz QCLs. Two configurations of cold fingers, i.e., type A and B with asymmetric and symmetric thermal dissipation schemes, respectively, are investigated here. A finite-element thermal analysis is carried out to study the parametric effects on the thermal management of the terahertz QCL. The modeling reveals that the symmetric thermal dissipation (type B) results in a more uniform thermal conduction and lower maximum temperature in the active region of the laser, compared to the traditional asymmetric thermal dissipation scheme (type A). To verify the simulation, experiments are further performed by measuring laser performance and comb characteristics of terahertz QCLs emitting around 4.2 THz mounted on type A and type B cold fingers. The experimental results show that the symmetric thermal dissipation approach (type B) is effective for improving the comb and dual-comb operation of terahertz QCLs, which can be further widely adopted for spectroscopy, imaging, and near-field applications.  
\end{abstract}

\section{Introduction}
Frequency combs\cite{fc3,fc2,fc1,mirfc,qclfc}, consisting of phase-coherent equidistant laser lines, have opened new avenues in many fields such as fundamental time metrology, imaging, communications, etc.\cite{ap2,imaging,ap3,ap1,FCS}. In particular, two frequency combs with slightly different repetition frequencies can be directly used for the high-precision spectroscopy by employing a multi-heterodyne dual-comb technique\cite{ap2}. In the terahertz regime, dual-comb spectroscopy has a number of advantages over conventional spectrometers, i.e., Fourier transform infrared (FTIR) and time domain spectroscopies (TDS), in the basic performance metrics of frequency resolution, accuracy, acquisition speed, signal to noise ratio (SNR), as well as in the potential for a compact system implementation\cite{fcscom1,DCS,acs,fcscom2}.

For a long time, frequency comb and dual-comb sources directly emitting in the terahertz frequency range have remained inexistent. Down-converted near-infrared frequency combs by nonlinear frequency conversion using photomixing or optical rectification have been widely used to generate terahertz frequency combs\cite{w1,w2}. However, the generated terahertz signal, especially the power per mode, is extremely weak, which results in problems for practical applications. The electrically-pumped terahertz quantum cascade lasers (QCLs) have rapidly become serious contenders to be operated as high power terahertz comb and dual-comb sources\cite{qcl3,qcl2,qcl4,qcl1}. A free running terahertz QCL can work in comb operation by exploiting the four-wave mixing locking mechanism of the nonlinear QCL gain medium. Furthermore, different approaches, i.e., group velocity dispersion engineering \cite{fork1984negative,burghoff2014terahertz,Faistoctave,villares2016dispersion,hillbrand2018tunable,zhou2019ridge}, active injection locking \cite{OE2010injection,barbieri2011coherent,li2015dynamics,hillbrand2019coherent} , and passive stabilization \cite{pl5} have been used to obtain broadband, highly stable frequency comb and dual-comb sources. Although the external active or passive stabilizations can produce more stable frequency comb and dual-comb operation, additional optical or electrical techniques are involved which make the comb system more complex. 

In this work, we investigate the thermal effects on stability of free running QCL frequency comb and dual-comb sources emitting around 4.2 THz. Two configurations of cold fingers with asymmetric (type A) and symmetric (type B) thermal dissipation schemes are investigated and then compared. Both simulation and experiments are performed for identical terahertz QCLs mounted on two different cold fingers. The results show that a symmetric heat dissipation scheme is in favor of the frequency comb and dual-comb operation. The measured inter-mode beatnote signal from the laser mounted on the type B cold finger in the dual-comb operation condition is 16 dB stronger than that measured from the same laser mounted on the type A cold finger. Furthermore, employing the symmetric thermal dissipation scheme, the free running dual-comb operation is observed up to 42 K, while the type A mounting configuration results in a maximum dual-comb operation temperature of 35 K. The improvement demonstrated in this work confirms that a better heat dissipation method can thermally affect the coherence and stability of comb/dual-comb operation, which can be adopted for stabilizing the frequency comb and dual-comb sources combining with other phase locking techniques, and then for practical applications.

\section{Design and simulations}

\begin{figure}[ht]
	\includegraphics[width=1\linewidth]{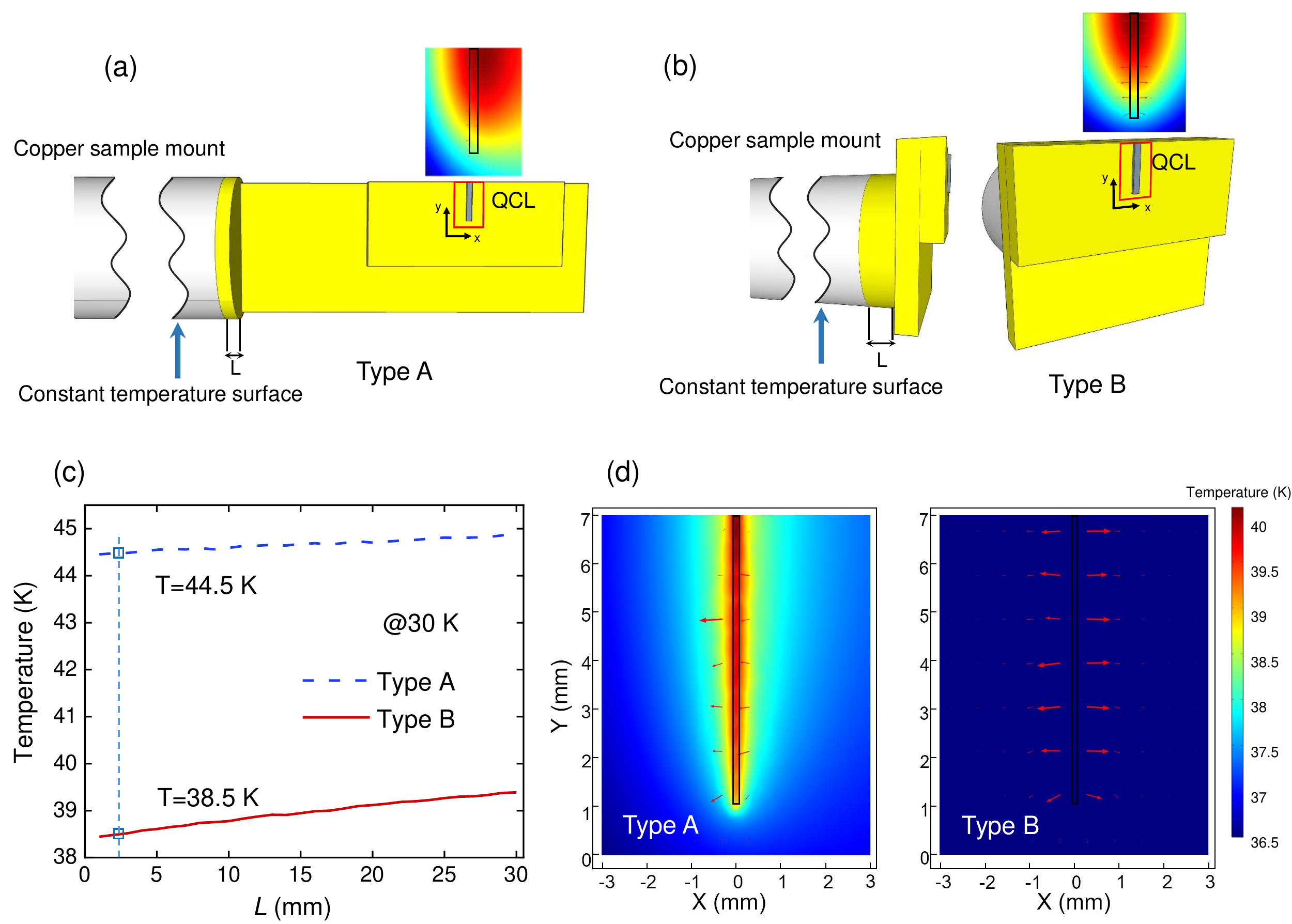}
	\caption{Schematics of type A (a) and type B (b) cold fingers. In (b), the left and right panels show the same cold finger viewed at different angles. $L$ denotes the thickness of the cylindrical parts of the two cold fingers, which are directly attached onto the copper sample mount of the cryostat. The figure insets in (a) and (b) show the calculated heat distributions on the interface planes, between the cold finger and the QCL heat sink, marked by the red rectangles. (c) Simulated temperature values in QCLs as a function of $L$ for type A (red) and type B (blue) cold fingers, respectively. Here, the temperature of the cold source (constant temperature surface) is fixed at 30 K. (d) Calculated heat distributions on the interface between the heat sink and the QCL chip for type A and type B cold fingers, respectively. The two graphs share the same color bar. Red arrows denote directions of the heat flux. For the simulation in (d), the temperature of the cold source is fixed at 30 K and $L$=2 mm.}
	\label{simulation}
\end{figure}

Figures \ref{simulation}(a) and \ref{simulation}(b) schematically show the configurations of type A and type B cold fingers, respectively. For the type A cold finger, the QCL device is mounted on the long arm of the cold finger and the thermal transfer is not symmetric [see Fig. \ref{simulation}(a)] because the cold source (constant temperature surface marked by a blue arrow) is on the left side of the QCL. The thickness of the arm or flat part of the type A holder, on which the QCL device is mounted, is 2.5 mm. While, in the newly designed type B cold finger, the cold source is under the QCL chip and a symmetric heat dissipation is expected.

To better understand the heat distributions of the two different cold finger configurations, a 3D finite-element simulation (COMSOL) based on the heat transfer theory is performed. For a steady-state problem, the heat transfer in solids can be described as the following equation
$$\nabla \cdot (\kappa\nabla T)+Q=0,\eqno(1)$$ 
where $\kappa$ is the thermal conductivity, $T$ is the temperature and $Q$ is the input power density. Thermal conductivities, imported directly from the material library of the COMSOL software, are listed as follows: 400 Wm$^{-1}$K$^{-1}$ for the heat sink and cold finger (copper); 46 Wm$^{-1}$K$^{-1}$ for the QCL device (GaAs). The QCL device is set as a heat source and the Joule heating power is determined by the electrical characteristics of the QCL [see Fig. \ref{LIV}]. When the QCL is operated at 1000 mA, the electrical power is around 5 W. By considering the dimensions of the QCL, i.e., 6 mm long, 150 $\mu$m wide and 150 $\mu$m thick, the thermal power density, $Q$, is calculated to be 1.35$\times$10$^{-10}$ Wm$^{-3}$. The temperature sensor is placed 0.4 inch away from the end of the copper sample mount. To simplify the simulation, this plane is set as a constant temperature surface as shown in Figs. \ref{simulation}(a) and \ref{simulation}(b). Convection and heat radiation are not considered in this simulation and we assume that thermal boundaries of the model are thermally isolated from the environment.

Figure \ref{simulation}(c) shows the simulated maximum temperature of the QCL device as a function $L$ [the length of the cylindrical parts as shown in Figs. \ref{simulation}(a) and \ref{simulation}(b)], obtained when the cold source is stabilized at 30 K. We choose 30 K because in the experiment our lasers can be safely stabilized at this temperature in the entire current dynamic range. For comb and dual-comb measurements, we normally stabilize the temperature at 30 K. Moreover, note that in this work, the lasers are cooled down using an open-flow liquid helium cryostat system to avoid vibration noises for comb operation. If we can operate the laser at a higher temperature, we can save the helium for an economical reason. For both cold finger configurations, one can see that the temperature increases with the increase of $L$. It can be understood that as $L$ is increased, the QCL device is positioned further away from the cold source (constant temperature surface), and therefore, the cooling effect is weakened. And it is expected that the temperature of the QCL device increases with $L$. In addition, as shown in Fig. \ref{simulation}(c), a difference between type A and type B cold fingers can be clearly observed, i.e., for all various $L$ values, the calculated QCL temperatures for the type B cold finger are always lower than those calculated for type A. The lower temperatures obtained for the type B cold finger are partially resulted from the symmetric thermal conduction as shown in the figure inset of Fig. \ref{simulation}(b). Specifically, for $L$=2 mm, the calculated temperature values of the QCL device mounted on type A and type B cold fingers are 44.5 K and 38.5 K, respectively, marked by squares in Fig. \ref{simulation}(c).

In Fig. \ref{simulation}(d), the calculated temperature distribution on the interface between the copper heat sink and the QCL chip, for both cold finger configurations, are shown with red arrows indicating the directions of the heat flux. It can be seen that as the temperature of the cold source is fixed at 30 K, in the same temperature scale, the simulated temperature around the QCL chip mounted on the type A cold finger is higher than that on the type B cold finger. Furthermore, a clear difference in the heat flux direction can be observed between the two different cold fingers. For the type A cold finger as shown in the left panel of Fig. \ref{simulation}(d), the red arrows on the left of the laser ridge are bigger than those on the right hand side, which indicates that the heat dissipates towards the left direction because the cold source is on its left [see Fig. \ref{simulation}(a)]. However, for the type B cold finger [right panel of Fig. \ref{simulation}(d)], a symmetric heat flux towards left and right directions is obtained, which facilitates a more uniform heat dissipation in the QCL device. For these reasons, the calculated maximum temperature in the QCL mounted on the type A cold finger is $\sim$5.9 K higher than that mounted on the type B cold finger. In Fig. S1 in Supplement 1, the calculated heat distribution of the type B cold finger in a smaller temperature scale is shown.

\section{Experimental results and discussion}

In Fig. \ref{simulation}, the thermal simulation shows that the newly designed type B cold finger can result in a lower temperature, and especially, an uniform heat dissipation in QCL chips. To verify the simulation, in this section, we experimentally compare the laser performance and comb operation of the same terahertz QCL mounted on two different cold fingers. To minimize the artificial effects as much as possible in the laser mounting process for a fair comparison between the two types of cold fingers, such as, nonuniform pressure applied onto the laser bar, misalignment between the laser bar and the edge of the copper heat sink, etc., a ``pick and place" machine was used for the laser mounting. The machine can softly ``pick" a single laser chip, place it at the wanted position with high precision, and then apply an uniform pressure onto the laser bar for the indium soldering.

\begin{figure}[!b]
	\centering\includegraphics[width=0.8\linewidth]{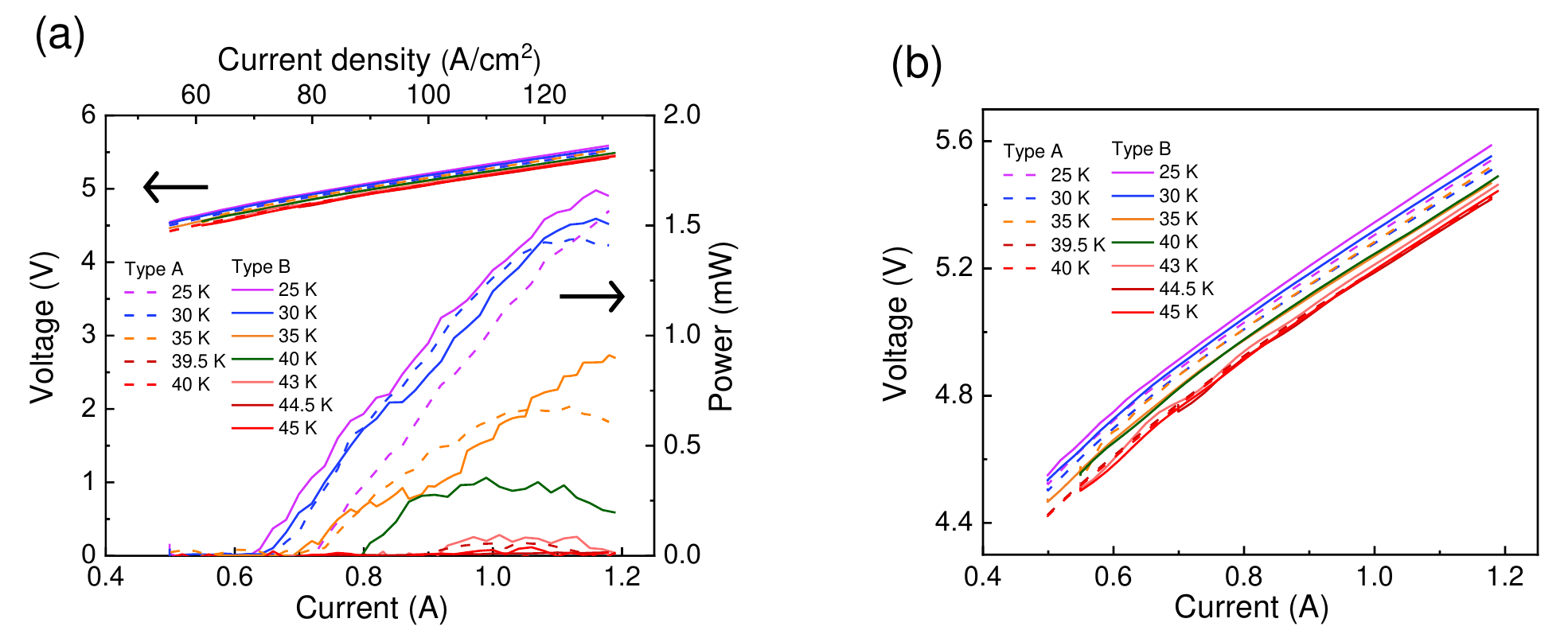}
	\caption{(a) Measured light-current-voltage characteristics in continuous wave (cw) mode of the QCL mounted on type A (dashed line) and type B (solid line) cold fingers. (b) Enlarged view of current-voltage characteristics shown in (a). The QCL ridge is 6 mm long and 150 $\mu$m wide.}
	\label {LIV}
\end{figure}

The terahertz QCL used in this work is based on a hybrid active region (bound-to-continuum for photon emission and resonant-phonon for fast carrier depopulation) emitting around 4.2 THz\cite{wan2017homogeneous}. The laser ridge is 6 mm long and 150 $\mu$m wide. The dimensions are identical to the ones used in the simulation. Figure \ref{LIV}(a) shows the light-current-voltage ($L-I-V$) characteristics of the QCL measured at different heat sink temperatures in continuous wave (CW) mode. Solid and dashed curves denote the measured results for the same QCL mounted on type B and type A cold fingers, respectively. In Fig. \ref{LIV}(b), we zoom in the $I-V$ curves shown in Fig. \ref{LIV}(a) to clearly show the small difference in electrical transport for the same laser mounted on two difference cold fingers. While, the $L-I$ curves indicate that at a same heat sink temperature, the laser mounted on the type B cold finger can produce slightly higher output power. For example, at 25 K, we obtain a power of 1.66 mW for type B versus a power of 1.56 mW for type A. In addition, the measured maximum operation temperature for the type B cold finger is 5 K higher than that for the type A cold finger, which agrees well with the simulation results. To prove the reproducibility of the results shown in Fig. \ref{LIV}, we warmed up the system, removed the QCL device from the cold finger, re-mount the laser, and performed $L-I-V$ measurements again. The results are shown in Fig. S2 (Supplement 1). It indicates that although the collected light power shown in Fig. S2 is slightly different from the results shown in Fig. 2 (due to the small difference in the optical alignment), the measured maximum operation temperature values for the same QCL device mounted on type A and type B cold fingers are 39 and 43 K, respectively, with a difference of 4 K. It can be seen that both results shown in Fig. \ref{LIV} and Fig. S2 agree well with each other.

\begin{figure}[!b]
	\includegraphics[width=1\linewidth]{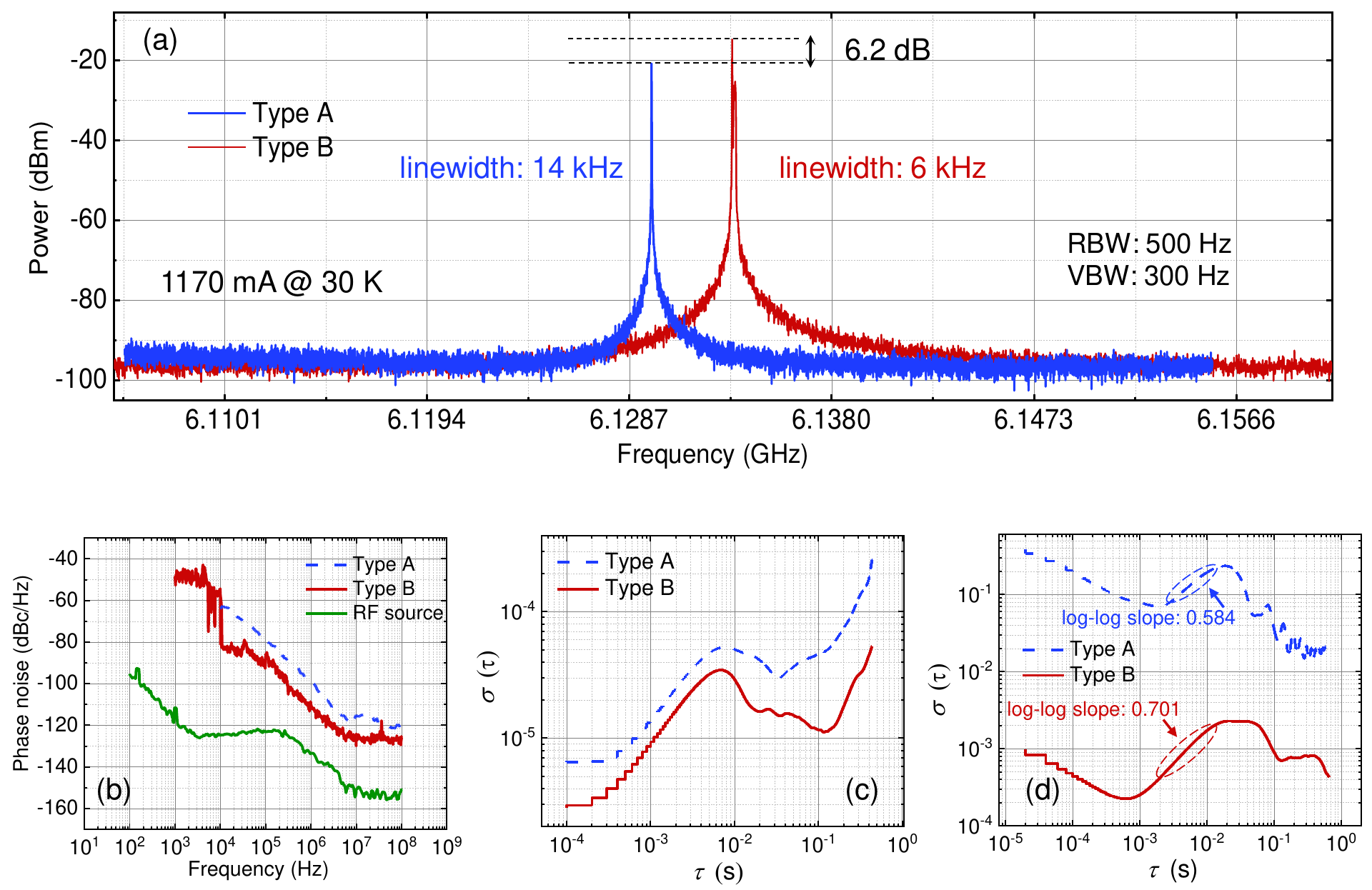}
	\caption{Inter-mode beatnote spectra and their stability evaluation of the QCL mounted on type A and type B cold fingers. (a) Measured inter-mode beatnote spectra of the QCL comb mounted on type A (blue) and type B (red) cold fingers. A resolution bandwidth (RBW) of 500 Hz is used for the measurement. (b), (c), and (d) are measured phase noise, frequency Allan deviation, and amplitude Allan deviation of the beatnote signals in (a), respectively. In (b), the phase noise is measured using a RBW of 10\%. As a reference, the phase noise of a signal at  6.13 GHz with a power of -14 dBm generated from a commercial signal generator (R\&S, SMA100B) is plotted as a green curve for comparison. In (c), the gate time used for the frequency Allan deviation measurements is 100 $\mu$s. In (d), the sweep time used for the amplitude Allan deviation measurements is 1 s. All the results are recorded when the QCL is operated at a drive current of 1170 mA at a heat sink temperature of 30 K.}
	\label {comb}
\end{figure}

In Fig. \ref{comb}, we present the effects of cold finger configurations on the QCL frequency comb operation. The QCL used for this measurement is the one reported in Fig. \ref{LIV}. The inter-mode beatnote, which is resulted from the beatings of the neighboring comb lines, has been widely used to characterize the stability of the repetition rate of a QCL comb. Because the QCL cavity is normally in a length of few millimeters, the inter-mode beatnote frequency is from few gigahertz to tens of gigahertz. And the electrical beatnote signal can be easily measured using the QCL itself as a detector by monitoring its current modulation introduced by the optical beating\cite{gellie2010injection,li2015dynamics}. Figure \ref{comb}(a) shows the recorded inter-mode beatnote spectra around 6.13 GHz of the same QCL mounted on type A (blue curve) and type B (red curve) cold fingers. The resolution bandwidth (RBW) and video bandwidth (VBW) parameters used for the measurement are 500 Hz and 300 Hz, respectively. Note that, without any external locking mechanisms, the measured linewidths for type A and type B cases are 14 kHz and 6 kHz, respectively. A signal to noise ratio (SNR) of 84 dB is obtained for the type B (symmetric thermal dissipation) cold finger, which is 6 dB larger than that measured for the type A (asymmetric thermal dissipation) cold finger.

In Figs. \ref{comb}(b), \ref{comb}(c), and \ref{comb}(d), we plot the measured phase noise, frequency Allan deviation, and amplitude Allan deviation of the beatnote signals shown in Fig. \ref{comb}(a). The phase noise was measured using the phase noise module of a spectrum analyzer. The Allan deviation is a classic statical method that has been widely used to evaluate the stability of a signal with time. For the frequency or amplitude Allan deviations, we first recorded frequency or amplitude values as a function of time, and then a formula \cite{allan1987,pl6} was used to mathematically calculate the Allan deviation plots. Overall, compared to the type A cold finger, the type B cold finger brings about improvements in the phase noise and Allan deviations. Specifically, the inter-mode beatnote phase noises measured at 1 MHz, for type A and type B cold fingers, are -98 dBc/Hz and -112 dBc/Hz, respectively, as shown in Fig. \ref{comb}(b). Note that, the phase noise for the type B cold finger can be measured starting from 1 kHz, while for the type A cold finger the minimum measurable frequency is 10 kHz. The smaller measurable frequency infers that the phase noise is improved by employing the symmetric heat dissipation. As a reference, the phase noise of a signal generated from a commercially available microwave generator (Rohde \& Schwarz, SMA100B) at 6.13 GHz with a power of -14 dBm is also plotted in Fig. \ref{comb}(b) for a comparison. In Fig. \ref{comb}(c), we report the frequency Allan deviation as a function of integration time, measured using a frequency counter with a gate time of 100 $\mu$s. Although an increase of the frequency Allan deviation with time can be observed, at each time the frequency Allan deviation measured for the type B cold finger is smaller than that measured for the type A cold finger. Similarly, the measured amplitude Allan deviations for the type B cold finger are also smaller than those measured for the type A cold finger as shown in Fig. \ref{comb}(d). It comes to a conclusion from the above results that a more well-balanced and symmetrical heat dissipation method can bing about an improvement in QCL comb performance. Therefore, the stability of inter-mode beatnote of the QCL comb can be significantly improved in terms of phase noise, frequency and amplitude Allan deviations. Note that for the same inter-mode beatnote signal, it is normal that the frequency and amplitude stabilities with time show different behaviors. From Figs. \ref{comb}(c) and \ref{comb}(d), it can be seen that in the time range studied here, the best stabilities for frequency and amplitude are obtained as the integration times are around 0.1 ms and 1 ms, respectively. Furthermore, for long $\tau$, the frequency and amplitude Allan deviations show opposite trend with time.

\begin{figure}[!t]
	\includegraphics[width=1\linewidth]{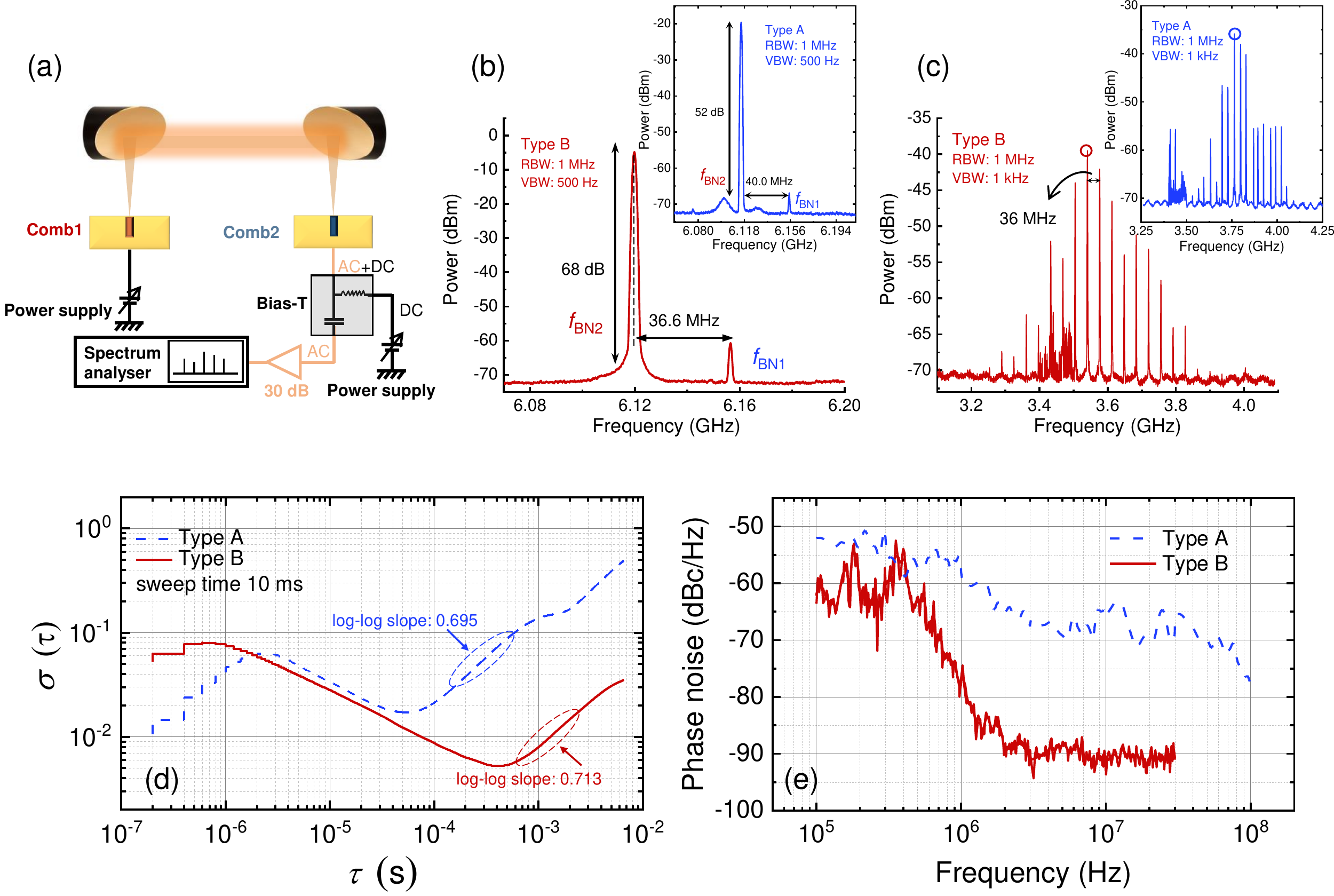}
	\caption{(a) Experimental setup of the terahertz dual-comb operation. Comb1 and Comb2 are terahertz QCLs with identical ridge dimensions, i.e., 6 mm long and 150 $\mu$m wide. The inter-mode beatnote and dual-comb signals are measured via Comb2 as a fast terahertz detector. (b) Inter-mode beatnote signals measured for the QCLs mounted on type B cold fingers in the dual-comb operation condition with a RBW of 1 MHz and a VBW of 500 Hz. $f\rm _{BN1}$ and $f\rm _{BN2}$ are the inter-mode beatnotes of Comb1 and Comb2, respectively. The inset is the corresponding results obtained from same lasers mounted on type A cold fingers. (c) Dual-comb spectrum around 3.6 GHz measured via Comb2 with a RBW of 1MHz and a VBW of 1 kHz when both QCLs are mounted on type B cold fingers. The inset is the dual-comb spectrum measured from same QCLs mounted on type A cold fingers. (d) Amplitude Allan deviation plots measured using a sweep time of 10 ms for dual-comb lines at 3.58 GHz [red circle in (c), type B cold finger] and 3.78 GHz [blue circle in the inset of (c), type A cold finger]. (e) Phase noise spectra of the dual-comb lines marked in (c). The red solid and blue dashed curves are for type B and type A cold fingers. For the dual-comb operation, Comb1 and Comb2 are operated at 1040 mA and 1170 mA, respectively. The heat sink temperature for both laser combs are stabilized at 30 K. }
	\label {dc}
\end{figure}

We further investigate the heat dissipation effect on the dual-comb performance. The experimental setup of the terahertz QCL dual-comb system is shown in Fig. \ref{dc}(a). Two terahertz QCLs with identical dimensions, i.e., 6 mm long and 150 $\mu$m wide, are mounted on type A and then type B cold fingers for comparisons. Two parabolic mirrors were used for the optical coupling between the two QCLs. The multiheterodyne dual-comb spectra were measured using the laser comb2 as a fast detector, and then amplified by 30 dB and directly registered on a spectrum analyser\cite{acs,pl6}. Figures \ref{dc}(b) and \ref{dc}(c) show the measured inter-mode beatnote and dual-comb spectra of the QCLs mounted on type B cold fingers, while the corresponding results measured for type A cold fingers are shown in the figure insets. It is worth noting that although both Figs. \ref{dc}(b) and \ref{comb}(a) show the inter-mode beatnote spectra, the SNR of the signals are different due to the change of operation conditions. In the dual-comb operation condition, due to the effect of the optical feedback, we can see the SNR of the inter-mode beatnote is smaller than that measured in the single comb operation condition for both type A and type B cold fingers. However, if we focus only onto the dual-comb operation situation, it can be seen from Fig. \ref{dc}(b) that the measured SNR for $f$$_\textrm{BN2}$ are 68 dB and 52 dB for type B and type A cold fingers, respectively. The 16 dB improvement in SNR induced by the deployment of type B cold finger is observed.

The frequency difference between $f$$_\textrm{BN1}$ and $f$$_\textrm{BN2}$, i.e., 40.0 MHz for type A cold finger and 36.6 MHz for type B cold finger, determines the line spacing of the down-converted dual-comb spectra. In Fig. \ref{dc}(c) and its inset, we report the dual-comb spectra for type B and type A cold fingers, respectively, measured with a RBW of 1 MHz and a VBW of 1 kHz. Although a significant improvement in inter-mode beatnote SNR for the type B cold finger can be clearly observed, we find that the SNRs of dual-comb spectra for both type A and type B cold fingers are almost identical. This can be explained as follows: the dual-comb is resulted from the multiheterodyne beatings of two sets of comb lines from two different QCLs. The dual-comb operation is not only related to the respective performance of two QCLs, but also associated with the optical coupling. A slight disturbance in optical coupling caused by mechanical vibrations and/or temperature fluctuations can massively affect the dual-comb performance. In this dual-comb experiment, the two laser combs are positioned in two different cryostats and they cannot share same noise fluctuations. In view of this, no significant difference in SNRs of dual-comb spectra for both type A and type B can be observed.

To investigate the stability or noise properties of the dual-comb lines, the amplitude Allan deviation and phase noise analysis of selected dual-comb lines for type A and type B cold fingers are carried out. Two dual-comb lines for two different cold finger configurations, i.e., 3.58 GHz for type B cold fingers (red circle) and 3.78 GHz for type A cold fingers (blue circle) as shown in Fig. \ref{dc}(c), are selected. Figure \ref{dc}(d) shows the amplitude Allan deviation plots of the two selected dual-comb lines. It can be clearly seen that although the SNRs of the dual-comb lines for type A and type B cold fingers are almost identical [see Fig. \ref{dc}(b)], the amplitude stability is different for the two dual-comb lines. From Fig. \ref{dc}(d), we can see that the amplitude Allan deviation first decreases and then increases with integration time. The dual-comb line from the type B mounting scheme shows longer amplitude stability than that for the type A scheme, i.e., 415 $\mu$s for type B versus 54 $\mu$s for type A. Furthermore, in Figs. \ref{dc}(d) and \ref{comb}(d), the log-log slopes for the integration time ranges indicated by the dashed ovals are displayed for comparison. Although the two figures show the amplitude Allan deviations for two different types of signals, i.e., Fig. \ref{comb}(d) for inter-mode beatnote and Fig. \ref{dc}(d) for dual-comb signals, the random walk trend (characterized by a log-log slope of 0.5)\cite{random-walk} can be observed for the two types of signals. Because the dual-comb signal is resulted from the multiheterodyne beatings of two laser combs, its long-term stability is much worse than the inter-mode beatnote signal (see Figs. \ref{comb} and \ref{dc}). If we consider similar integration time ranges (see Fig. S3 in Supplement 1 for longer integration times), the amplitude Allan deviations for inter-mode beatnote and dual-comb signals show similar trends with each other. The phase noise spectra of the two dual-comb lines are shown in Fig. \ref{dc}(e). A lower phase noise can be obtained for the type B cold finger (see the red curve). At 1 MHz, the measured phase noise for type B and type A cold fingers are -77 and -58 dBc/Hz, respectively. Note that for the amplitude Allan deviation and phase noise measurements of the dual-comb lines, two different measurements taken for same devices and identical operation conditions can be largely different if the two measurements are performed on different days. Therefore, in this work, to perform a fair comparison, the data shown in Figs. \ref{comb} and \ref{dc} are obtained on the same day and the time duration between the two measurements for type A and type B cold fingers is as short as possible. To prove the reproducibility that the type B cold finger is more favorable for the stable dual-comb operation, in Figs. S3 and S4 (Supplement 1), we report two other dual-comb measurements taken on different days. The experimental results indicate that for all dual-comb measurements taken on different days, the type B cold finger always results in higher amplitude stability and lower phase noise (although the specific Allan deviation and phase noise values are not exactly identical). It is worth noting that in Fig. S4 (Supplement 1), the heat sink temperature for the type B measurement is stabilized at 35 K that is 5 K higher than the temperature for the type A measurement. Even though, we still observe the improved dual-comb operation for the type B scheme. In principle, this result can refer that the improved performance is resulted from the symmetric heat dissipation of type B cold finger rather than the 5-K difference in the operation temperature between the type A and type B mounting schemes (see Figs. \ref{simulation} and \ref{LIV}).

\begin{figure}[h]
	\centering\includegraphics[width=0.7\linewidth]{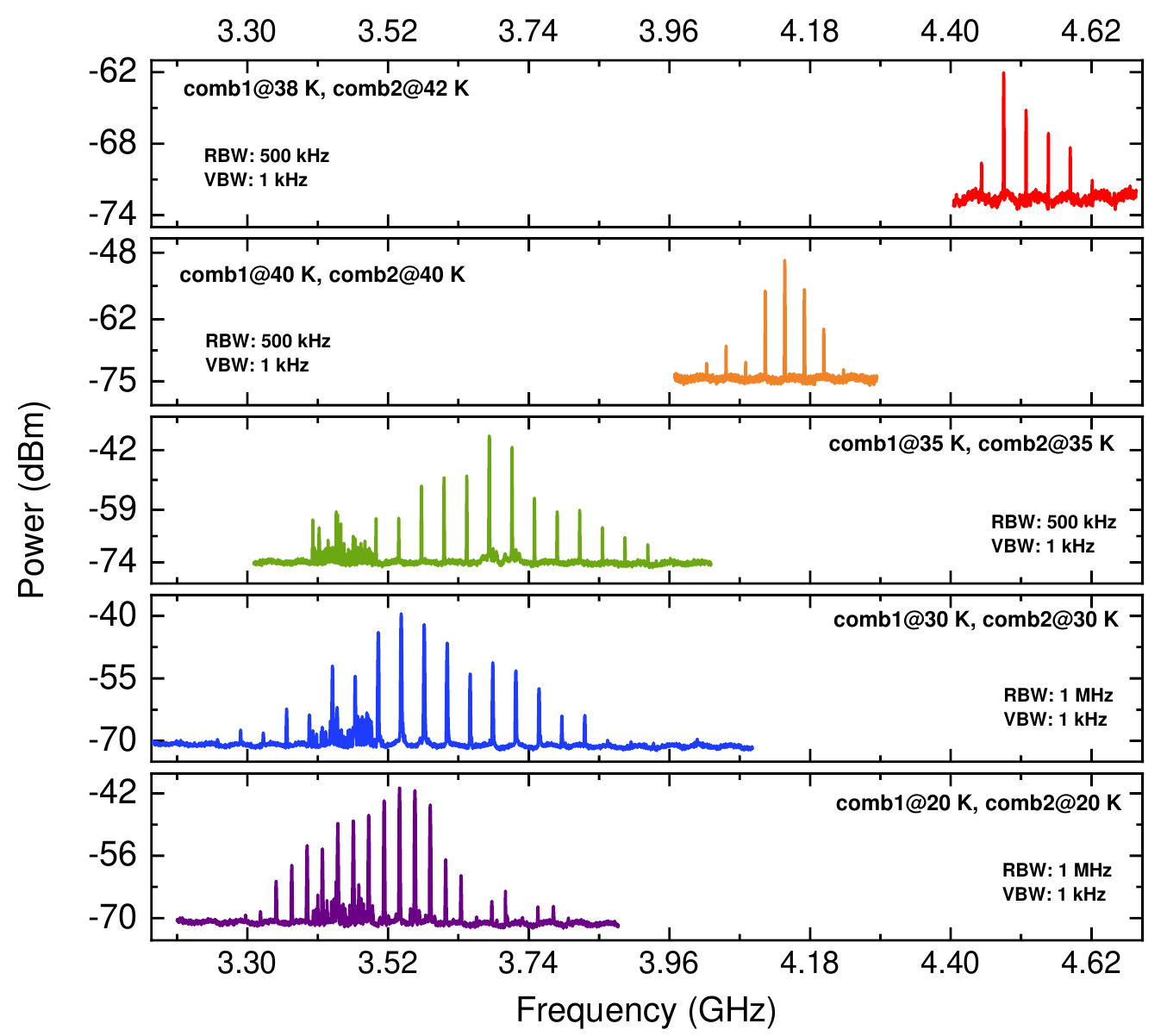}
	\caption{Dual-comb spectra recorded at different temperatures with the QCL combs mounted on type B cold fingers. From bottom to top panels, the operation temperature is increased.}
	\label {temp}
\end{figure}

In Figs. \ref{LIV}-\ref{dc}, we experimentally show that the symmetric thermal dissipation scheme (type B cold finger) is beneficial to the laser performance, frequency comb and dual-comb operation. In Fig. \ref{temp}, we present a study on the temperature performance of the dual-comb operation of QCLs mounted on type B cold fingers. From bottom to top panels, we show dual-comb spectra recorded at different temperatures from 20 K to 42 K. As temperature is increased, the dual-comb line number and power of the lines decrease. It is because the power and number of the terahertz comb lines are decreasing with temperature. Because the frequency comb and dual-comb operation are extremely sensitive to temperature, we can see the central frequencies and line spacing of the measured dual-comb spectra change with temperature. Although the dual-comb line number decreases with temperature, the dual-comb operation with 6 comb lines is successfully observed up to 42 K as shown in the top panel of Fig. \ref{temp}. Note that for type A cold fingers, because the maximum laser operation temperature is 40 K (see Fig. \ref{LIV}), the maximum temperature for dual-comb operation is below 40 K and in a range of 35-38 K.

Note that in a recent publication \cite{ForrerAPL2021}, the frequency comb operation of a four-well terahertz QCL with the low-loss copper-copper double-metal waveguide was observed up to 80 K which is much higher than the highest operation temperature reported in this work (40 K). We have to say that in this work we focus on the improved frequency comb and dual-comb operation resulted from the use of the symmetric thermal dissipation (type B cold finger). Here the active region and waveguide of the terahertz QCL were not optimized for high operation temperature. The value of the study here is that the relative advantages by using the cold finger type B (symmetric thermal distribution) is generally correct and independent on the specific laser device itself. Besides the 5-K increase in the maximum operation temperature, it can be seen from Figs. 3 and 4 that the coherence and stability of the comb and dual-comb operation have been dramatically improved by using the type B cold finger. Moreover, although our lasers show lower operation temperature, the inter-mode beatnote demonstrates much larger SNR, i.e., 84 dB (measured with a RBW of 500 Hz) in this work versus 45 dB (measured with a RBW of 300 Hz) in \cite{ForrerAPL2021}.

\section{Conclusion}

In conclusion, we have demonstrated that a symmetric heat dissipation scheme can give rise to an improvement in the laser performance, frequency comb and dual-comb operation of terahertz QCLs emitting around 4.2 THz. A 3D finite-element thermal simulation was firstly performed to calculate the temperature distributions and directions of heat flux for two different cold fingers (asymmetric versus symmetric heat dissipation). Experimental results show that, compared to the asymmetric cold finger, the symmetric heat dissipation scheme can result in a 5-K increase in the maximum laser operation temperature, a 16-dB stronger inter-mode beatnote, more stabilized dual-comb spectra, and higher operation temperatures for the dual-comb operation. Both experiment and simulation show good agreements with each other. Such proposed symmetric thermal dissipation method shows the practical use for improvement of comb and dual-comb operation of terahertz QCLs, which can be further utilized combining with traditional phase locking techniques to obtain more stable comb and dual-comb sources for spectroscopy, imaging, and near-field applications.

\section*{Funding}

National Natural Science Foundation of China (62022084, 61875220, 62035005, 61927813, 61704181, and 61991432); Ministry of Science and Technology of the People's Republic of China (2017YFF0106302); Chinese Academy of Sciences (ZDBS-LY-JSC009 and YJKYYQ20200032); Science and Technology Commission of Shanghai Municipality (20XD1424700).
	
\section*{Disclosures}

The authors declare no conflicts of interest. 
	
\section*{Data Availability Statement}
	Data underlying the results presented in this paper are not publicly available at this time but may be obtained from the authors upon reasonable request.

\section*{Supplemental document}
    See Supplement 1 for supporting content.

\bibliography{references}

\begin{thebibliography}{10}
\newcommand{\enquote}[1]{``#1''}

\bibitem{fc3}
T.~Udem, R.~Holzwarth, and T.~Hansch, \enquote{{Optical frequency metrology},}
  {\protect\JournalTitle{{Nature}}} \textbf{{416}}, {233--237} ({2002}).

\bibitem{fc2}
T.~Yasui, Y.~Kabetani, E.~Saneyoshi, S.~Yokoyama, and T.~Araki,
  \enquote{{Terahertz frequency comb by multifrequency-heterodyning
  photoconductive detection for high-accuracy, high-resolution terahertz
  spectroscopy},} {\protect\JournalTitle{{Appl. Phys. Lett.}}} \textbf{{88}},
  241104 ({2006}).

\bibitem{fc1}
P.~Del'Haye, A.~Schliesser, O.~Arcizet, T.~Wilken, R.~Holzwarth, and T.~J.
  Kippenberg, \enquote{{Optical frequency comb generation from a monolithic
  microresonator},} {\protect\JournalTitle{{Nature}}} \textbf{{450}},
  {1214--1217} ({2007}).

\bibitem{mirfc}
A.~Schliesser, N.~Picque, and T.~W. Haensch, \enquote{{Mid-infrared frequency
  combs},} {\protect\JournalTitle{{Nat. Photonics}}} \textbf{{6}}, {440--449}
  ({2012}).

\bibitem{qclfc}
J.~Faist, G.~Villares, G.~Scalari, M.~Rosch, C.~Bonzon, A.~Hugi, and M.~Beck,
  \enquote{{Quantum Cascade Laser Frequency Combs},}
  {\protect\JournalTitle{{Nanophotonics}}} \textbf{{5}}, {272--291} ({2016}).

\bibitem{ap2}
F.~Keilmann, C.~Gohle, and R.~Holzwarth, \enquote{{Time-domain mid-infrared
  frequency-comb spectrometer},} {\protect\JournalTitle{{Opt. Lett.}}}
  \textbf{{29}}, {1542--1544} ({2004}).

\bibitem{imaging}
T.~Ideguchi, S.~Holzner, B.~Bernhardt, G.~Guelachvili, N.~Picque, and T.~W.
  Haensch, \enquote{{Coherent Raman spectro-imaging with laser frequency
  combs},} {\protect\JournalTitle{{Nature}}} \textbf{{502}}, {355--358}
  ({2013}).

\bibitem{ap3}
C.~G. Wade, N.~Sibalic, N.~R. de~Melo, J.~M. Kondo, C.~S. Adams, and K.~J.
  Weatherill, \enquote{{Real-time near-field terahertz imaging with atomic
  optical fluorescence},} {\protect\JournalTitle{{Nat. Photonics}}}
  \textbf{{11}}, {40--43} ({2017}).

\bibitem{ap1}
F.~Cappelli, L.~Consolino, G.~Campo, I.~Galli, D.~Mazzotti, A.~Campa, M.~S.
  de~Cumis, P.~C. Pastor, R.~Eramo, M.~Rosch, M.~Beck, G.~Scalari, J.~Faist,
  P.~De~Natale, and S.~Bartalini, \enquote{{Retrieval of phase relation and
  emission profile of quantum cascade laser frequency combs},}
  {\protect\JournalTitle{{Nat. Photonics}}} \textbf{{13}}, {562--568} ({2019}).

\bibitem{FCS}
N.~Picque and T.~W. Haensch, \enquote{{Frequency comb spectroscopy},}
  {\protect\JournalTitle{{Nat. Photonics}}} \textbf{{13}}, {146--157} ({2019}).

\bibitem{fcscom1}
G.~Villares, A.~Hugi, S.~Blaser, and J.~Faist, \enquote{{Dual-comb spectroscopy
  based on quantum-cascade-laser frequency combs},}
  {\protect\JournalTitle{{Nat. Commun.}}} \textbf{{5}}, 1--9 ({2014}).

\bibitem{DCS}
I.~Coddington, N.~Newbury, and W.~Swann, \enquote{{Dual-comb spectroscopy},}
  {\protect\JournalTitle{{Optica}}} \textbf{{3}}, {414--426} ({2016}).

\bibitem{acs}
H.~Li, Z.~Li, W.~Wan, K.~Zhou, X.~Liao, S.~Yang, C.~Wang, J.~Cao, and H.~Zeng,
  \enquote{Toward compact and real-time terahertz dual-comb spectroscopy
  employing a self-detection scheme,} {\protect\JournalTitle{ACS Photonics}}
  \textbf{7}, 49--56 (2020).

\bibitem{fcscom2}
L.~A. Sterczewski, J.~Westberg, Y.~Yang, D.~Burghoff, J.~Reno, Q.~Hu, and
  G.~Wysocki, \enquote{{Terahertz Spectroscopy of Gas Mixtures with Dual
  Quantum Cascade Laser Frequency Combs},} {\protect\JournalTitle{{ACS
  Photonics}}} \textbf{{7}}, {1082--1087} ({2020}).

\bibitem{w1}
I.~A. Finneran, J.~T. Good, D.~B. Holland, P.~B. Carroll, M.~A. Allodi, and
  G.~A. Blake, \enquote{Decade-spanning high-precision terahertz frequency
  comb,} {\protect\JournalTitle{Phys. Rev. Lett.}} \textbf{114}, 163902 (2015).

\bibitem{w2}
T.~Yasui, R.~Ichikawa, Y.-D. Hsieh, K.~Hayashi, H.~Cahyadi, F.~Hindle,
  Y.~Sakaguchi, T.~Iwata, Y.~Mizutani, H.~Yamamoto, K.~Minoshima, and H.~Inaba,
  \enquote{Adaptive sampling dual terahertz comb spectroscopy using dual
  free-running femtosecond lasers,} {\protect\JournalTitle{Sci. Rep.}}
  \textbf{5}, 10786 (2015).

\bibitem{qcl3}
R.~Kohler, A.~Tredicucci, F.~Beltram, H.~Beere, E.~Linfield, A.~Davies,
  D.~Ritchie, R.~Iotti, and F.~Rossi, \enquote{{Terahertz
  semiconductor-heterostructure laser},} {\protect\JournalTitle{{Nature}}}
  \textbf{{417}}, {156--159} ({2002}).

\bibitem{qcl2}
S.~Barbieri, M.~Ravaro, P.~Gellie, G.~Santarelli, C.~Manquest, C.~Sirtori,
  S.~P. Khanna, E.~H. Linfield, and A.~G. Davies, \enquote{{Coherent sampling
  of active mode-locked terahertz quantum cascade lasers and frequency
  synthesis},} {\protect\JournalTitle{{Nat. Photonics}}} \textbf{{5}},
  {306--313} ({2011}).

\bibitem{qcl4}
M.~Brandstetter, C.~Deutsch, M.~Krall, H.~Detz, D.~C. MacFarland,
  T.~Zederbauer, A.~M. Andrews, W.~Schrenk, G.~Strasser, and K.~Unterrainer,
  \enquote{{High power terahertz quantum cascade lasers with symmetric wafer
  bonded active regions},} {\protect\JournalTitle{{Appl. Phys. Lett.}}}
  \textbf{{103}}, 171113 ({2013}).

\bibitem{qcl1}
M.~Roesch, G.~Scalari, M.~Beck, and J.~Faist, \enquote{{Octave-spanning
  semiconductor laser},} {\protect\JournalTitle{{Nat. Photonics}}}
  \textbf{{9}}, {42--47} ({2015}).

\bibitem{fork1984negative}
R.~Fork, O.~Martinez, and J.~Gordon, \enquote{Negative dispersion using pairs
  of prisms,} {\protect\JournalTitle{Opt. Lett.}} \textbf{9}, 150--152 (1984).

\bibitem{burghoff2014terahertz}
D.~Burghoff, T.-Y. Kao, N.~Han, C.~W.~I. Chan, X.~Cai, Y.~Yang, D.~J. Hayton,
  J.-R. Gao, J.~L. Reno, and Q.~Hu, \enquote{Terahertz laser frequency combs,}
  {\protect\JournalTitle{Nat. Photonics}} \textbf{8}, 462 (2014).

\bibitem{Faistoctave}
M.~R{\"o}sch, G.~Scalari, M.~Beck, and J.~Faist, \enquote{Octave-spanning
  semiconductor laser,} {\protect\JournalTitle{Nat. Photonics}} \textbf{9},
  42--47 (2015).

\bibitem{villares2016dispersion}
G.~Villares, S.~Riedi, J.~Wolf, D.~Kazakov, M.~J. S{\"u}ess, P.~Jouy, M.~Beck,
  and J.~Faist, \enquote{Dispersion engineering of quantum cascade laser
  frequency combs,} {\protect\JournalTitle{Optica}} \textbf{3}, 252--258
  (2016).

\bibitem{hillbrand2018tunable}
J.~Hillbrand, P.~Jouy, M.~Beck, and J.~Faist, \enquote{Tunable dispersion
  compensation of quantum cascade laser frequency combs,}
  {\protect\JournalTitle{Opt. Lett.}} \textbf{43}, 1746--1749 (2018).

\bibitem{zhou2019ridge}
K.~Zhou, H.~Li, W.~Wan, Z.~Li, X.~Liao, and J.~Cao, \enquote{Ridge width effect
  on comb operation in terahertz quantum cascade lasers,}
  {\protect\JournalTitle{Appl. Phys. Lett.}} \textbf{114}, 191106 (2019).

\bibitem{OE2010injection}
P.~Gellie, S.~Barbieri, J.-F. Lampin, P.~Filloux, C.~Manquest, C.~Sirtori,
  I.~Sagnes, S.~P. Khanna, E.~H. Linfield, A.~G. Davies, H.~Beere, and
  D.~Ritchie, \enquote{Injection-locking of terahertz quantum cascade lasers up
  to 35ghz using rf amplitude modulation,} {\protect\JournalTitle{Opt.
  Express}} \textbf{18}, 20799--20816 (2010).

\bibitem{barbieri2011coherent}
S.~Barbieri, M.~Ravaro, P.~Gellie, G.~Santarelli, C.~Manquest, C.~Sirtori,
  S.~P. Khanna, E.~H. Linfield, and A.~G. Davies, \enquote{Coherent sampling of
  active mode-locked terahertz quantum cascade lasers and frequency synthesis,}
  {\protect\JournalTitle{Nat. Photonics}} \textbf{5}, 306 (2011).

\bibitem{li2015dynamics}
H.~Li, P.~Laffaille, D.~Gacemi, M.~Apfel, C.~Sirtori, J.~Leonardon,
  G.~Santarelli, M.~R{\"o}sch, G.~Scalari, M.~Beck, J.~Faist, W.~H\"{a}nsel,
  R.~Holzwarth, and S.~Barbieri, \enquote{Dynamics of ultra-broadband terahertz
  quantum cascade lasers for comb operation,} {\protect\JournalTitle{Opt.
  Express}} \textbf{23}, 33270--33294 (2015).

\bibitem{hillbrand2019coherent}
J.~Hillbrand, A.~M. Andrews, H.~Detz, G.~Strasser, and B.~Schwarz,
  \enquote{Coherent injection locking of quantum cascade laser frequency
  combs,} {\protect\JournalTitle{Nature Photonics}} \textbf{13}, 101--104
  (2019).

\bibitem{pl5}
H.~Li, M.~Yan, W.~Wan, T.~Zhou, K.~Zhou, Z.~Li, J.~Cao, Q.~Yu, K.~Zhang, M.~Li,
  J.~Nan, B.~He, and H.~Zeng, \enquote{Graphene-coupled terahertz semiconductor
  lasers for enhanced passive frequency comb operation,}
  {\protect\JournalTitle{Advanced Science}} \textbf{6}, 1900460 (2019).

\bibitem{wan2017homogeneous}
W.~Wan, H.~Li, T.~Zhou, and J.~Cao, \enquote{Homogeneous spectral spanning of
  terahertz semiconductor lasers with radio frequency modulation,}
  {\protect\JournalTitle{Sci. Rep.}} \textbf{7}, 44109 (2017).

\bibitem{gellie2010injection}
P.~Gellie, S.~Barbieri, J.-F. Lampin, P.~Filloux, C.~Manquest, C.~Sirtori,
  I.~Sagnes, S.~P. Khanna, E.~H. Linfield, A.~G. Davies, H.~Beere, and
  D.~Ritchie, \enquote{Injection-locking of terahertz quantum cascade lasers up
  to 35ghz using rf amplitude modulation,} {\protect\JournalTitle{Opt.
  Express}} \textbf{18}, 20799--20816 (2010).

\bibitem{allan1987}
D.~W. Allan, \enquote{Should classical variance be used as a basic measure in
  standards metrology?} {\protect\JournalTitle{IEEE Trans. Instrum. Meas.}}
  \textbf{36}, 646--54 (1987).

\bibitem{pl6}
Y.~Zhao, Z.~Li, K.~Zhou, X.~Liao, W.~Guan, W.~Wan, S.~Yang, J.~C. Cao, D.~Xu,
  S.~Barbieri, and H.~Li, \enquote{{Active Stabilization of Terahertz
  Semiconductor Dual-Comb Laser Sources Employing a Phase Locking Technique},}
  {\protect\JournalTitle{Laser Photonics Rev.}} \textbf{15}, 2000498 (2021).

\bibitem{random-walk}
D.~V. Land, A.~P. Levick, and J.~W. Hand, \enquote{The use of the allan
  deviation for the measurement of the noise and drift performance of microwave
  radiometers,} {\protect\JournalTitle{Measurement Science and Technology}}
  \textbf{18}, 1917--1928 (2007).

\bibitem{ForrerAPL2021}
A.~Forrer, Y.~Wang, M.~Beck, A.~Belyanin, J.~Faist, and G.~Scalari,
  \enquote{Self-starting harmonic comb emission in thz quantum cascade lasers,}
  {\protect\JournalTitle{Appl. Phys. Lett.}} \textbf{118}, 131112 (2021).

\end{thebibliography}






\end{document}